\documentclass[pre,aps,floatfix,amsmath,twocolumn]{revtex4-1}

\usepackage{amsmath}
\usepackage{epsfig}
\usepackage{array}
\usepackage{verbatim}
\usepackage{hyperref}
\usepackage{amssymb}
\usepackage{graphicx}

\usepackage{color}

\newcommand{\lref}[2][]{\hyperref[#2]{#1~\ref*{#2}}}
\renewcommand{\eqref}[1]{\hyperref[#1]{(\ref*{#1})}}

\numberwithin{equation}{section}

\newcommand{\mc}[1]{\mathcal{#1}}

\newcommand{\nn}{\nonumber}
\newcommand{\la}{\langle}
\newcommand{\ra}{\rangle}

\def\be{\begin{equation}}
\def\ee{\end{equation}}
\def\bea{\begin{eqnarray}}
\def\eea{\end{eqnarray}}

\newcommand{\eqa}[1]{\begin{align}#1\end{align}}

\newcommand{\RN}[1]{
	\textup{\uppercase\expandafter{\romannumeral#1}}%
}
	
\usepackage{color}
\definecolor{dgreen}{rgb}{0,0.7,0}

\usepackage{url,hyperref}

\usepackage{verbatim}
\usepackage{graphicx,epstopdf}  

\usepackage{amsfonts}
\usepackage{color}
\usepackage{dcolumn}   
\usepackage{bm}        
\usepackage{amssymb}   
\usepackage{hyperref}
\usepackage{dcolumn}   
\usepackage{bm} 
\usepackage{amssymb}  
\usepackage{verbatim}
\usepackage{breqn}
\usepackage{array}
\usepackage{enumerate}

\def\be{\begin{equation}}
\def\ee{\end{equation}}
\def\bea{\begin{eqnarray}}
\def\eea{\end{eqnarray}}
\def\bed{\begin{dmath*}}
	\def\eed{\end{dmath*}}

\graphicspath{{./}}
\newcommand{\beg}{\begin{equation}}

\newcommand{\en}{\end{equation}}

\renewcommand{\emph}{\textit}

\newcommand{\beq}{\begin{equation}}

\newcommand{\eeq}{\end{equation}}
\newcommand{\barray}{\begin{eqnarray}}
\newcommand{\earray}{\end{eqnarray}}

\def\be{\begin{equation}}
\def\ee{\end{equation}}
\def\bea{\begin{eqnarray}}
\def\eea{\end{eqnarray}}

\begin{document}
    
    \title{Transport properties of the classical Toda chain: effect of  a pinning potential}
    
    \author{Abhishek Dhar$^1$, Aritra Kundu$^1$, Joel L. Lebowitz$^{2,3}$, Jasen A. Scaramazza$^3$}
    \date{\today}                            
    
    \affiliation{$^1$ International Centre for Theoretical Sciences - Tata Institute of Fundamental Research$,$ Bengaluru$,$  560089 $,$ India \\$^2$ Department of Mathematics$,$  Rutgers University$,$  Piscataway$,$  NJ 08854$,$  USA\\ $^3$ Department of Physics and Astronomy$,$  Rutgers University$,$  Piscataway$,$ NJ 08854$,$  USA}

    \begin{abstract}
        We consider energy transport in the classical Toda chain in the presence of an additional pinning potential. The pinning potential is expected to destroy the integrability of the system and an interesting question is to see the signatures of this breaking of integrability on energy transport.  We investigate this by a study of the non-equilibrium steady state of the system connected to heat baths as well as the study of equilibrium correlations. Typical signatures of integrable systems are a  size-independent energy current, a flat bulk temperature profile and ballistic scaling of equilibrium dynamical correlations, these results being valid in the thermodynamic limit. We find that, as expected, these properties change drastically on introducing the pinning potential in the Toda model. 
        In particular, we find that the effect of a harmonic pinning potential is drastically smaller at low temperatures, compared to a quartic pinning potential. We explain this by noting that at low temperatures the Toda potential can be approximated by a harmonic inter-particle potential for which the addition of harmonic pinning does not destroy integrability.
    \end{abstract}
    
    \date{\today}
    \maketitle
    \section{Introduction}
    The transport of thermal energy in Hamiltonian systems is a problem of great theoretical and practical interest [\onlinecite{Lepri,BLR,LLP2003,dhar2008}]. In its simplest form, one considers heat flow in the non-equilibrium stationary state (NESS) of a system in contact with two thermal reservoirs at different temperatures. Very little is known rigorously about this problem except in the case of harmonic crystals [\onlinecite{RLL}] or hard rods in 1D [\onlinecite{spohn}]. These models are special cases of the larger class of integrable models, whose extensive numbers of conserved quantities are expected in general to lead to ballistic heat transport [\onlinecite{zotos,mazur,suzuki}]. This means that if a system of length $N$ (and cross-section $A$) is put in contact with heat reservoirs at temperatures $T_L$ and $T_R$, $T_L > T_R$, at its left and right ends, then the heat flow in the stationary state $J$ would be (except for boundary effects) independent of $N$. Another distinctive feature of the NESS of integrable systems is the flat temperature profile observed in the bulk of the system. These features
    are indeed observed for the harmonic chain and the hard particle gas and also for other integrable models such as the Toda lattice [\onlinecite{toda2,shyo,zotos,casati2014}].
    
    In contrast, one finds a different picture for generic  nonlinear non-integrable systems  such as Fermi-Pasta-Ulam chains [\onlinecite{Lepri1997,mai2007,Zhao2012,das2013}], the diatomic Toda chain [\onlinecite{hatano}] and the alternate mass hard particle gas [\onlinecite{dhar2001,grass2002,casati}]  where, simulations and various phenomenological theories find instead $J \sim N^{-\alpha}$ with $0< \alpha < 1$.  This appears to be the case for momentum conserving systems and this is referred to as anomalous transport. When a non-integrable system does not conserve momentum, for example, due to pinning by a one body potential, the transport is generally expected to be diffusive, also called ``normal'' since it satisfies Fourier's law, with $\alpha = 1$ and this has been seen in many simulations [\onlinecite{dhar2008}]. The temperature profiles observed in non-integrable models is also completely different from the flat ones in integrable models, here one finds instead that the temperature changes gradually from the hot end to the cold end.

    Apart from the non-equilibrium setups, signatures of non-integrability/integrability and anomalous transport also manifest themselves in the form of dynamical equilibrium spatiotemporal correlation functions. 
    In fact, significant theoretical progress in understanding anomalous transport in momentum conserving systems has been obtained by using the framework of nonlinear fluctuating hydrodynamics   [\onlinecite{Narayan2002,Beijerin2012,mendl2013,Spohn2013}], which allows one to make specific predictions for the form of equilibrium correlations of conserved quantities. Using ideas of linear response theory, one can then relate anomalous features in equilibrium correlations to those observed in the non-equilibrium set-up. For integrable systems, with a large number of conserved quantities, there is much ongoing work to develop a hydrodynamic framework [\onlinecite{doyon2018}]. One expects 'ballistic scaling' of correlation functions for integrable systems and this was observed in recent numerical work on the Toda chain [\onlinecite{kundu2016}]. A surprising exception to this expectation is the recent observation of ballistic scaling, but also anomalous and diffusive scaling of correlations in different parameter regimes of the integrable $XXZ$ model [\onlinecite{prosen2013}].
    
    An interesting question is the effect of adding extra terms to an integrable Hamiltonian which generically one expects should make the system non-integrable. Several studies have addressed this question.  
    The pinned Toda system was studied in [\onlinecite{zhang}] where it was found that energy transfer to high-frequency modes is slow and energy equipartition is not observed in the studied time scale. 
    Similar features were observed in the trapped hard rod system  [\onlinecite{Cao2018}], where it was observed that the system become chaotic after a characteristic time scale but fails to thermalize even at extremely large times. 
    For momentum-conserving systems surprising features (e.g apparent diffusive transport) has been reported when a system is taken slightly out of integrability, 
    for example in the Fermi-Pasta-Ulam chain at low temperatures [\onlinecite{Zhao2012,das2013}] and the alternate mass hard particle gas, at a mass ratio close to one [\onlinecite{casati2014}].

    The present work addresses the effect of adding a pinning potential to the integrable Toda lattice, that is expected to make the system non-integrable. The system should then become diffusive and we study how this crossover takes place, by studying the size-dependence of current and the temperature profile in the NESS as well as the form of equilibrium dynamical correlation functions.  
    It has been observed in a few recent studies that depending on the form of the integrability-breaking term and other parameters such as temperature, the crossover can occur at extremely large system sizes.  Following some preliminary results by two of the present authors [\onlinecite{jasen2018}],  this was further investigated in [\onlinecite{DiCintio2018}] which demonstrated that the crossover to diffusive transport in the quadratically pinned Toda chain occurs at very large length scales. This was then attributed to the fact that solitons are the main energy carriers in this system and the quadratic pinning potential affects them rather weakly. In the present study, we confirm the finding in  [\onlinecite{DiCintio2018}] but propose a somewhat different understanding for the slow crossover. We show that at low temperatures and small system sizes, the harmonically pinned Toda chain, in fact, behaves like a pinned harmonic chain, and more so at strong pinning. This then also provides some understanding as to why a quartic pinning leads to a much faster approach to the diffusive regime.

    The plan of the paper is as follows:
    In Sec.~\ref{sec:modelnsetups} we define the model and the various setups
    that we use to study transport properties of the system. In Sec.~\ref{sec:Numericalresults:NESS} we discuss simulation results for the NESS for the quadratic pinned Toda chain and the quartic pinned Toda chain, and also present comparisons with a related pinned harmonic chain. In Sec.~\ref{sec:Numericalresults:Equilibrium} we present results for equilibrium dynamical correlation functions, while in  Sec.~\ref{sec:Numericalresults:Chaos} we compute the Lyapunov exponent for various cases and try to see possible connections of the slow transition to diffusion with chaotic properties of the system. Finally, we summarise and discuss our findings in Sec.~\ref{sec:summary}.

    \section{The model: Setups and some background}
    \label{sec:modelnsetups}
    
    The model  we consider is a $1$-dimensional chain of $N$  particles with positions $\{q_i\}$ and momenta $\{p_i\}$ for $i=1,\ldots,N$, described by the  classical Hamiltonian:
    \begin{align}
    H &= \sum_{i=1}^{N}\left[\frac{p_i^2}{2} + \frac{\nu^2}{z}q_i^z \right] +\sum_{i=0}^N
    V(q_{i+1}-q_i), \\
    &{\rm where}~~V(r)=\frac{a}{b}\exp(-b r),\nonumber
    \label{Ham}
    \end{align}
    the constants $a,b,\nu > 0$, while $z$ is taken to be an even positive integer.
    For $\nu=0$, the system is the usual Toda chain [\onlinecite{toda}], which is a well-known integrable model for both periodic and fixed boundary conditions [\onlinecite{henon,flaschka}]. 
    
    Unless otherwise specified, $V(r)$ will refer to the Toda interaction for the remainder of this work.
    Although the purely Toda potential is integrable, an addition of on-site  potential, i.e., $\nu \ne 0$ and $z = 2,4$ is expected to break the integrability of the $\nu = 0$ system when the number of particles is greater than $2$. Indeed, the only obvious conserved quantities when $\nu \ne 0$ are $H$ itself and the centre of mass term $h_{c}$
    \beg
    h_{c} = \frac{1}{2}\bigg(\sum_{i=0}^{N+1} p_i\bigg)^2 + \frac{\nu^2}{2}\bigg(\sum_{i=0}^{N+1} q_i \bigg)^2.
    \label{c}
    \en

    We study transport properties of the system using the following probes.

    (i) \textbf{Properties in the NESS} --- In this setup, the system is connected to two heat reservoirs at the boundaries:
    We take fixed boundary conditions $q_0=0,~q_{N+1}=0$ and couple the particles $1$ and $N$ of the chain to Langevin baths with a coupling constant $\mu$, which act as thermal reservoirs at temperatures $T_L$ and $T_R$ and induce a non-equilibrium steady state (NESS). The equations of motion are now given by
    \eqa{
        \dot{q}_i &= p_i, ~~1\le i\le N \\  
        \dot{p}_1 &= V'(q_{2}-q_1) - V'(q_1-q_0) - \nu^2  q_1^{z-1} - \mu p_1 + \eta_L, \nn \\
        \dot{p}_i &= V'(q_{i+1}-q_{i}) - V'(q_i-q_{i-1}) - \nu^2  q_i^{z-1},~ 2\le i\le N-1, \nn\\
        \dot{p}_N &= V'(q_{N+1}-q_N) - V'(q_N-q_{N-1}) - \nu^2  q_{N}^{z-1}- \mu p_N + \eta_R,~ \nn
        \label{eq:open}
    }
    where $V'(r) \equiv dV(r)/dr$ and $\eta_L,\eta_R$ are white Gaussian noise terms with zero mean and variance $ \langle \eta_L(t) \eta_L(t') \rangle =2\mu k_B T_L \delta(t-t')$, $ \langle \eta_R(t) \eta_R(t') \rangle =2\mu k_B T_R \delta(t-t')$.   In this setup, the central quantity of interest are the bulk temperature ($T_i = \la p_i^2\ra,~ i \in [1,N]$) and the average heat current $J$, which in the NESS is constant in time and equal to any of the following quantities
    \eqa{
        J_i & = -\bigg\langle\frac{1}{2}(p_i+p_{i+1})V'(q_{i+1}-q_i)\bigg\rangle ~i \in [2,N-1],\nn\\
        J_L &= \mu(T_L - \langle p_1^2\rangle),\\
        J_R &= \mu(\langle p_N^2 \rangle- T_R). 
        \label{NESSJ}}
    where $\langle\cdot\rangle$ refers to the NESS average, which in simulations is computed by first allowing the system sufficient time to relax to the NESS before time averaging. 
    In  this work, we will present results for the system-size-dependence of $J$ and the form of the temperature profile $T_i$ in the NESS for the pinned potentials with the two cases $z=2$ and $z=4$.

    (ii)\textbf{Dynamical correlation functions in thermal equilibrium}---
    The isolated set-up is used to study the spatiotemporal decay of equilibrium correlations. In this set-up, one usually considers a periodic ring with  the $N$  particles evolving with the Hamiltonian equations of motion 
    \eqa{
        \dot{q}_i&=p_i ~,  \nn \\
        \dot{p}_i&=  V'(q_{i+1}-q_{i}) - V'(q_i-q_{i-1})  - \nu^2  q_i^{z-1} ~, 
        \label{eqbeom}
    }
    for $i=1,\ldots,N$ and 
    with the periodic boundary conditions $q_{N+i}=q_i$.  The local energy is defined as $e_i=\frac{1}{2}p_i^2 +\frac{\nu^2}{z}q_i^z+ V(q_{i+1} - q_i)$. The Hamiltonian dynamics exactly conserves the total energy, $\sum_{i=1}^N e_i$. We then study the equilibrium spatio-temporal correlations of energy fluctuations in this system given by 
    \begin{equation}
    C(k,t) = \frac{1}{N} \sum_i [\la e(i+k,t) e(i,0) \ra - \la e \ra^2]~,
    \label{eqcorr}
    \end{equation}
    where we have used translation invariance of the problem to sum over sites. The spatio-temporal nature of the spread of these correlations give us information about the nature of underlying dynamics.  It is expected that at large times, this will have  the scaling form $C(k,t) = t^{-1/2} f(k/t^{1/2})$ for diffusive systems while for integrable systems one expects the ``ballistic'' scaling form $C(k,t) = t^{-1} f(k/t)$.  For systems with anomalous transport, this function has a more complicated structure [\onlinecite{Spohn2013}].
    In our simulations, the system is first equilibrated by attaching Langevin heat baths (at the same temperature $T$) to all sites of the lattice. After the system reaches equilibrium we draw random samples to create our equilibrium initial conditions which are then evolved with the  Hamiltonian dynamics. We average over many such initial conditions to compute the equilibrium correlation functions in Eq.~(\ref{eqcorr}).
    
    (iii){\bf Probing chaotic properties from the Lyapunov exponent}--- The Lyapunov exponent studies how a small fluctuation grows in time. For a Hamiltonian system  evolving with  Eq.~(\ref{eqbeom}), the following quantity can be used to quantify chaos in the system:
    \eqa{
        \lambda(t) =  \frac{1}{2N t} \sum_{i=1}^N  \langle \ln \left[ \frac{\partial q_i(t)}{\partial q_1(0)}\right]^2 \rangle~,
    }
    where the average $\langle ...\rangle$ denotes an average over initial conditions ${\bf q,p}$ chosen from the equilibrium distribution. 
    To evaluate this, consider a localised infinitesimal perturbation of a specified initial condition ${\bf q,p}$ given by $\delta q_1(t=0)=Q_1(t=0)$. Let the resulting change in the trajectory at time by given by $\delta q_i(t)=Q_i(t)$ and $\delta p_i(t)= P_i(t)$ for $i=1,2,\ldots,N$. 
    The equations satisfied by ${\bf Q,P}$ are given by
    \eqa{
        \dot{Q}_i &= P_i,     \label{eq:nnlequation} \\
        \dot{P}_i &= -{\nu}^2 (z-1) q_i^{z-2} Q_i  + V''(q_{i+1}(t)-q_{i}(t))(Q_{i+1} -Q_i) \nn\\ &- V''(q_{i}(t)-q_{i-1}(t))(Q_{i} -Q_{i-1}).\nn
    }
    We solve these equations along with Eqs.~(\ref{eqbeom}), with the initial conditions ${\bf q}(0),{\bf p}(0)$ chosen from the equilibrium distribution and  $Q_i(0) = \delta_{i,1},~P_i(0) = 0$ for all $i=1,2,\ldots,N$. Our quantity of interest is then given by
    \eqa{
        \lambda(t) =  \frac{1}{2N t} \sum_{i=1}^N  \langle \ln Q_i^2(t) \rangle~.
        \label{eq:lambdaD}
    }
    At large times, this gives the maximum Lyapunov exponent of the system  \eqa{\lambda = \lim\limits_{t\to \infty} \lambda(t).\label{eq:lyuponov}} 
    For an integrable system, it can be shown that the Lyapunov exponent  vanishes while for non-integrable systems they are expected to be positive.
    
    In the following sections, we will numerically study the above characteristics for three different models.
    
    \section{Simulation results for the NESS}
    \label{sec:Numericalresults:NESS}
    We study the non-equilibrium properties of (a) the quadratically pinned Toda chain at low and high temperatures and (b) the quartic pinned Toda chain. We also show that the results for case (a) of the quadratically pinned Toda lattice are similar to those obtained for a pinned harmonic lattice. 
    
    \textit{Details of simulations:}  In  the Langevin simulations with heat baths at two ends of the chain, the dynamics are integrated with a Brownian velocity-Verlet  algorithm [\onlinecite{allen}] with a time step of $dt \leq 0.005$. The system is first to let to run for time $\sim 10^7$ during which it reaches the steady state. Then statistics for temperature and current in steady state is collected for the next $\sim 10^7$ times with a gap of $10$ units.

    \begin{figure}
        \centering
        \includegraphics[width=\linewidth]{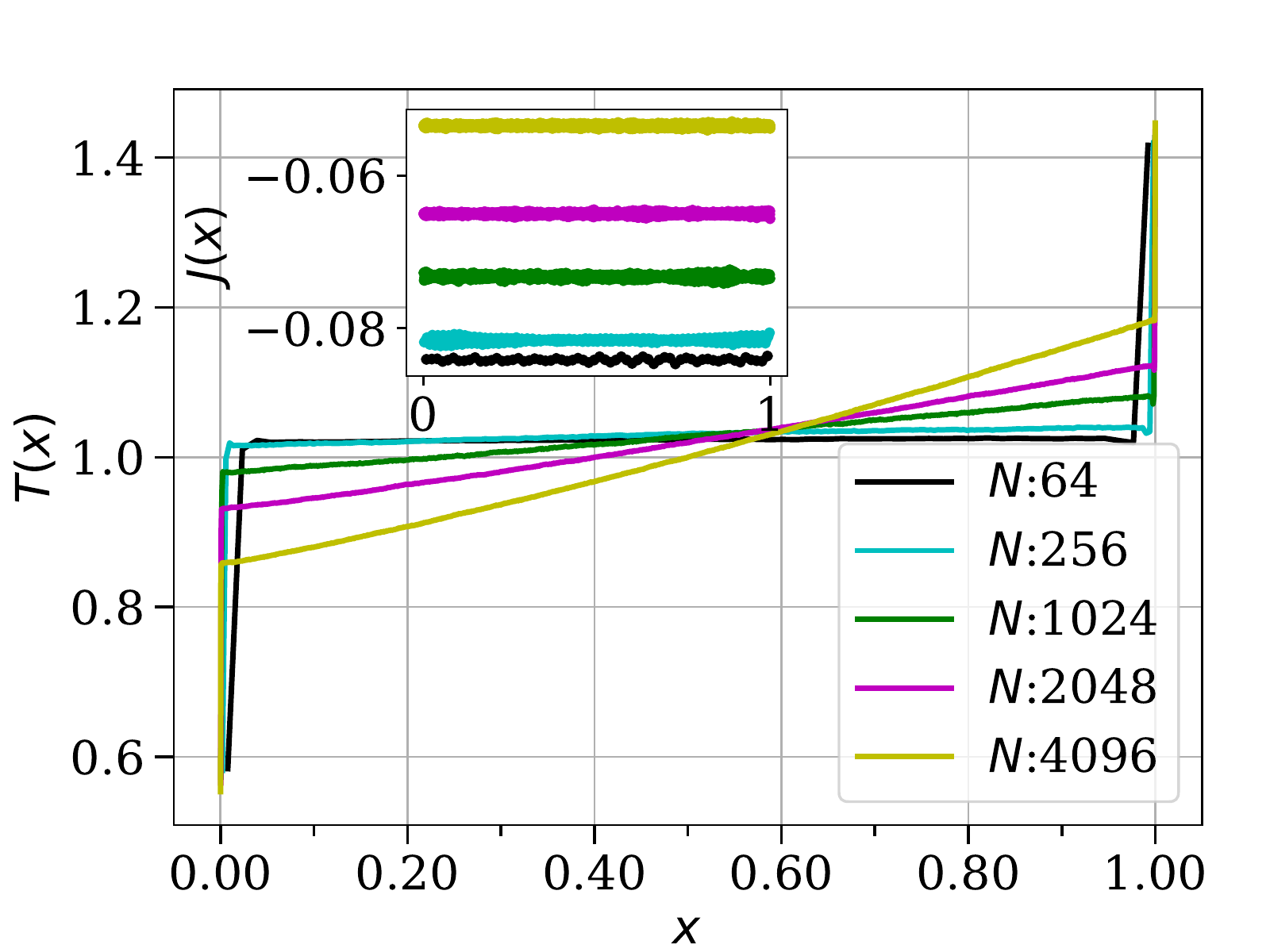}
        \caption{ Temperature profiles ($T$-vs $x=i/L$) for the Toda chain with harmonic pinning (Parameters $a = b  = 1$, $\nu=2$, $z = 2$, $T_L = 0.5$, $T_R = 1.5$). The temperature gradient is almost negligible at small system size which is bit unusual, but at larger system size it develops a small gradient. Also the current (shown in inset) is not independent of system size as one would expect in a ballistic, however it changes very slowly with system size. }
        \label{harmpin_N_LT}
    \end{figure}
    
    \begin{figure}
        \centering
        \includegraphics[width=\linewidth]{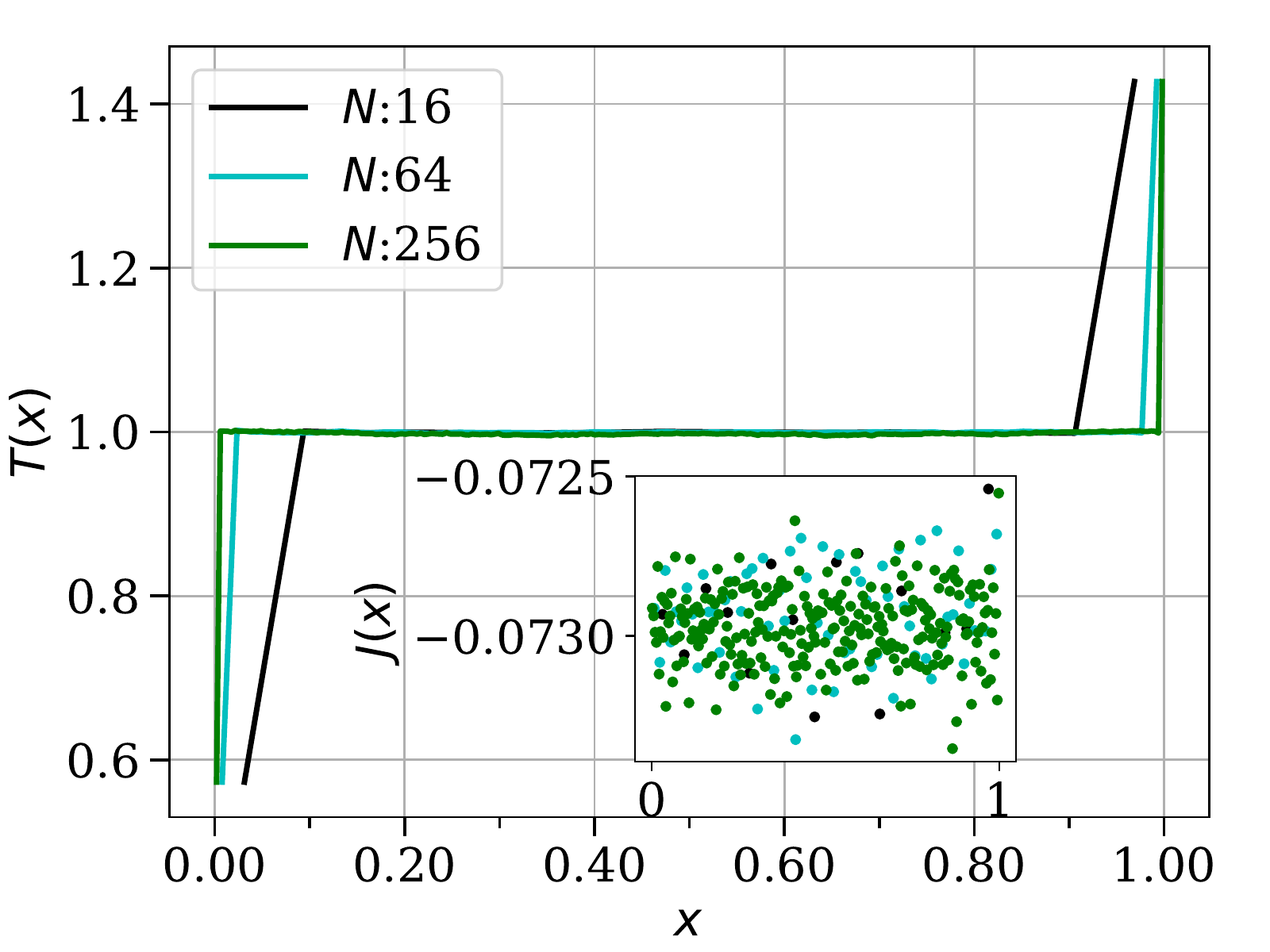}
        \caption{ Temperature profiles for the Harmonic chain with harmonic pinning ($a = b  = 1$, $\nu=2$, $z = 2$, $T_L = 0.5$, $T_R = 1.5$). The temperature is flat and there is no gradient across the system. Also the current (shown in inset) is independent with system size as $N^{0}$, which shows this has ballistic transport. }
        \label{harmpin_Har_LT}
    \end{figure}
    \textbf{(a) Quadratically pinned Toda chain:} This corresponds to the case  $z=2$ in Eq.~\ref{eq:open}. \\
    \textit{Low temperature:}
    The transport in this system shows unusual behaviour at low temperatures. In Fig.~\ref{harmpin_N_LT} we show the temperature and current profile at steady state at low temperature (with $T_L=0.5,~ T_R=1.5$) . For small system size, the temperature profile is almost flat, which might make one think of ballistic transport.  However, as noted in [\onlinecite{DiCintio2018}] this is not true and a slight temperature gradient develops at larger system sizes with the gradient being extremely small along with large boundary jumps. This is surprising and reminiscent of a diffusive system with an extremely long mean free path. As we will discuss now, this low-temperature behaviour arises because the system is in some sense close to the limit of a harmonic chain. 
    \\
    
\begin{figure}
        \centering
        \includegraphics[width=0.85\linewidth]{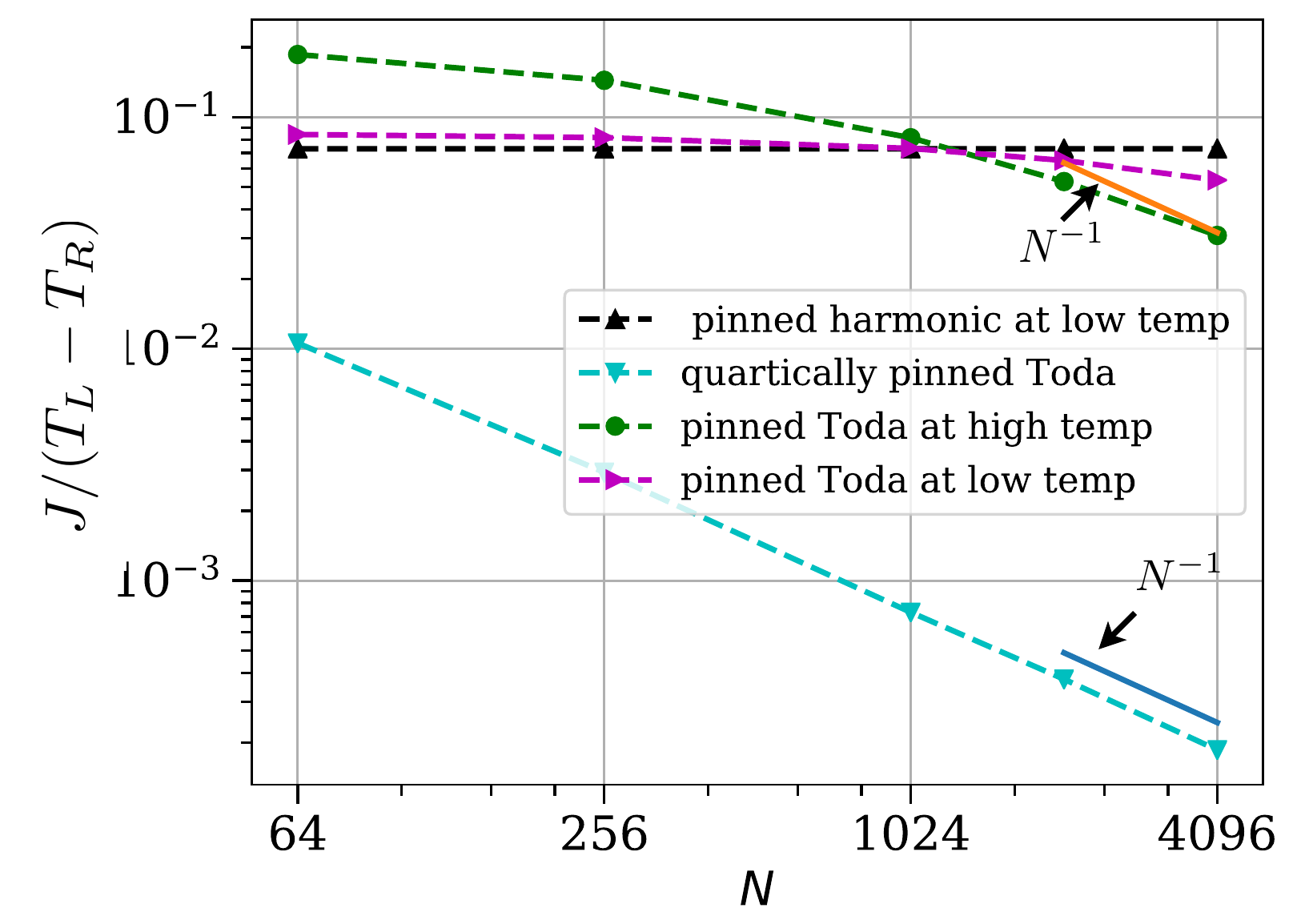}
        \caption{Current scaling with system size for various setups}
        \label{fig:currscaling}
    \end{figure}

    In Fig.~\ref{harmpin_Har_LT} we show the temperature and current profiles in the quadratically pinned harmonic  chain with the inter-particle coupling term $V(r) = \frac{1}{2} r^2$ and $z=2$ at low temperature (with $T_L = 0.5,~T_R=1.5$).
    This system is known to be integrable and we see expected features of an integrable system: A flat temperature profile as well as a current independent of system size. The value of the current, in this case, can be computed explicitly using the expressions given in [\onlinecite{rodh}] and the simulation results show very good agreement. 
    For completeness, we quote the formula for NESS current (for $N \to \infty$)   in the harmonic case with $V(r) = \frac{k}{2}r^2$ and $z=2$ as
    \eqa{J = \frac{T_L-T_R}{4\mu^3}\left(\mc{A}+ 2k\mu^2 - \sqrt{\mc{A}(\mc{A} + 4k\mu^2)}\right),}
    where $\mc{A} = k^2 + \nu^2 \mu^2$.
    We propose that at the low temperatures, the quadratically pinned Toda chain [ case (a)] is in some sense close to the quadratically pinned Harmonic chain. Due to pinning of the potential, at low enough temperatures, the lattice vibrations are very small and the effect of non-linearity of the Toda potential does not play a significant role. In Table.~(\ref{tab:comp}), we show the average bulk temperature and the current in the Toda chain for $N=32$ and see that at the strongest pinning case, the bulk temperature differs from the harmonic case by about $0.5 \%$ and the current by about $4\%$. The system size scaling of the current is shown in Fig.~(\ref{fig:currscaling}). 
    
\begin{figure}
        \centering
        \includegraphics[width=\linewidth]{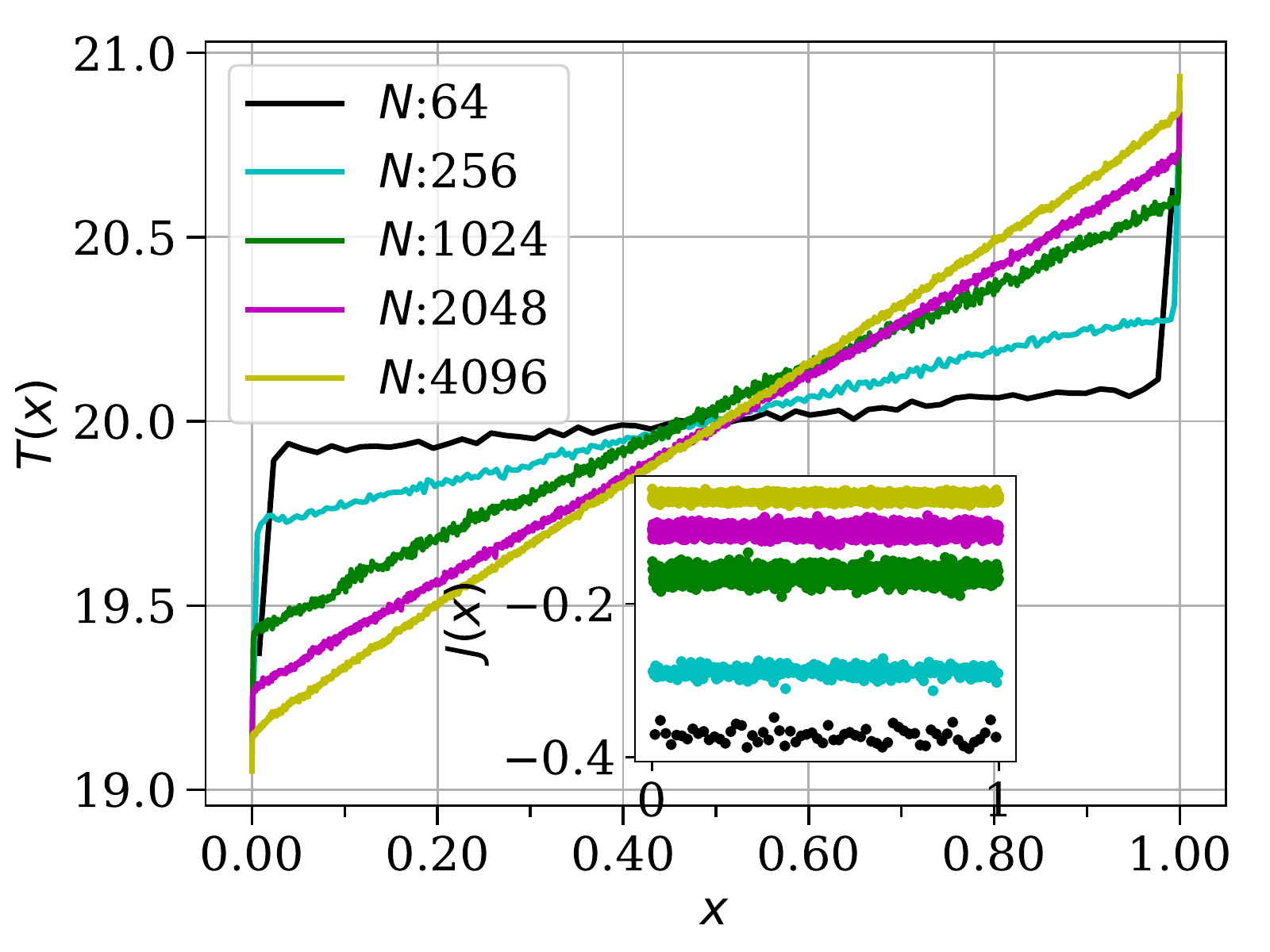}
        \caption{ Temperature profiles  for the Toda chain with harmonic pinning ($a = b  = 1$, $\nu=2$, $z = 2$, $T_L = 19$, $T_R = 21$). At high temperatures, there is a noticeable temperature gradient as we increase the system size. Also the current (shown in inset) decreases with system size as $\sim 1/N$. }
        \label{harmpin_N_HT}
    \end{figure}

    \begin{table}
        \begin{tabular}{|l|l|l|l|}
            \hline
            $    \nu$    &  $T_b$&  $J^{Toda}_{b}$&  $J^{Harmonic}_{b}$  \\ \hline
            1    & 1.012 & -0.1504  &-0.133974  \\ \hline
            2    & 1.019 & -0.0846 &-0.072949  \\ \hline
            4    & 1.012 &  -0.0287&-0.026389  \\ \hline
            6    &1.005  & -0.0134 &-0.012829  \\ \hline
        \end{tabular}
        \caption{Comparison of bulk temperature defined as $T_b = \sum_{i=2}^{N-1} T_i/(N-2)$ and bulk current $J_b =\sum_{i=2}^{N-1} J_i/(N-2)$  for pinned Toda chain and corresponding pinned Harmonic chain for system size $N=32$, $T_L=0.5$ and $T_R=1.5$. At this system size the bulk temperature for the harmonic case is very close to $T=1.0$, as can also be seen in Fig.~(\ref{harmpin_Har_LT}).}
        \label{tab:comp}
    \end{table}
    \textit{High temperature:}
    At high temperatures (with $T_L=19,~T_R=21$) as shown in Fig.~(\ref{harmpin_N_HT}), we find that the diffusive nature is more prominent and we find  a much larger temperature  gradient and correspondingly  smaller boundary jumps at the two ends. The current now shows a significant decay with system size but as seen in  Fig.~(\ref{fig:currscaling}), we are still not in the diffusive $ J \sim 1/N$ regime. This is consistent with the fact that at the largest system size we still see boundary jumps in the temperature profile. We also note that at the smallest system size $N=64$, the quantity $J/(T_L-T_R)$ differs from the expected   value for the harmonic chain (which is independent of temperature). In fact this value is much larger than the harmonic chain value, which is an indication that, in the high temperature case, the pinned Toda chain is far from the harmonic limit.

\begin{figure}
        \includegraphics[width=\linewidth]{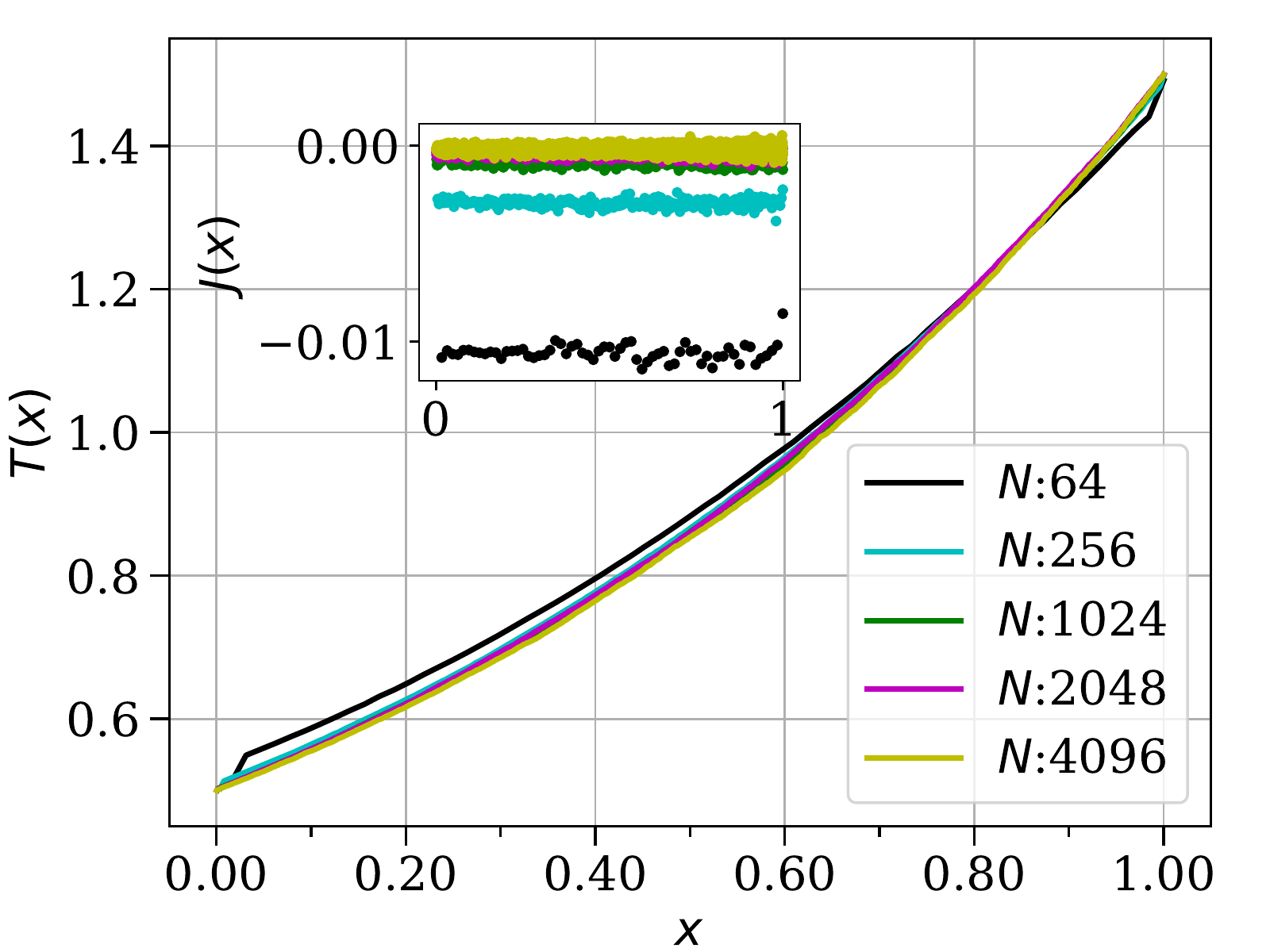}
        
        \caption{Temperature profiles and currents for the Toda chain with quartic pinning ($z = 4$), with all other parameters the same as in Fig.~\ref{harmpin_N_LT}. Note that with increasing $N$ the profile approaches a smooth curve between $T_L = 0.5$ and $T_R = 1.5$. Inset: Averaged site-by-site current profile, with the average current decaying with system size as $\sim 1/N$. }
        \label{quarpin_N}
    \end{figure}

\begin{figure}
	\centering
	\includegraphics[width=0.85\linewidth]{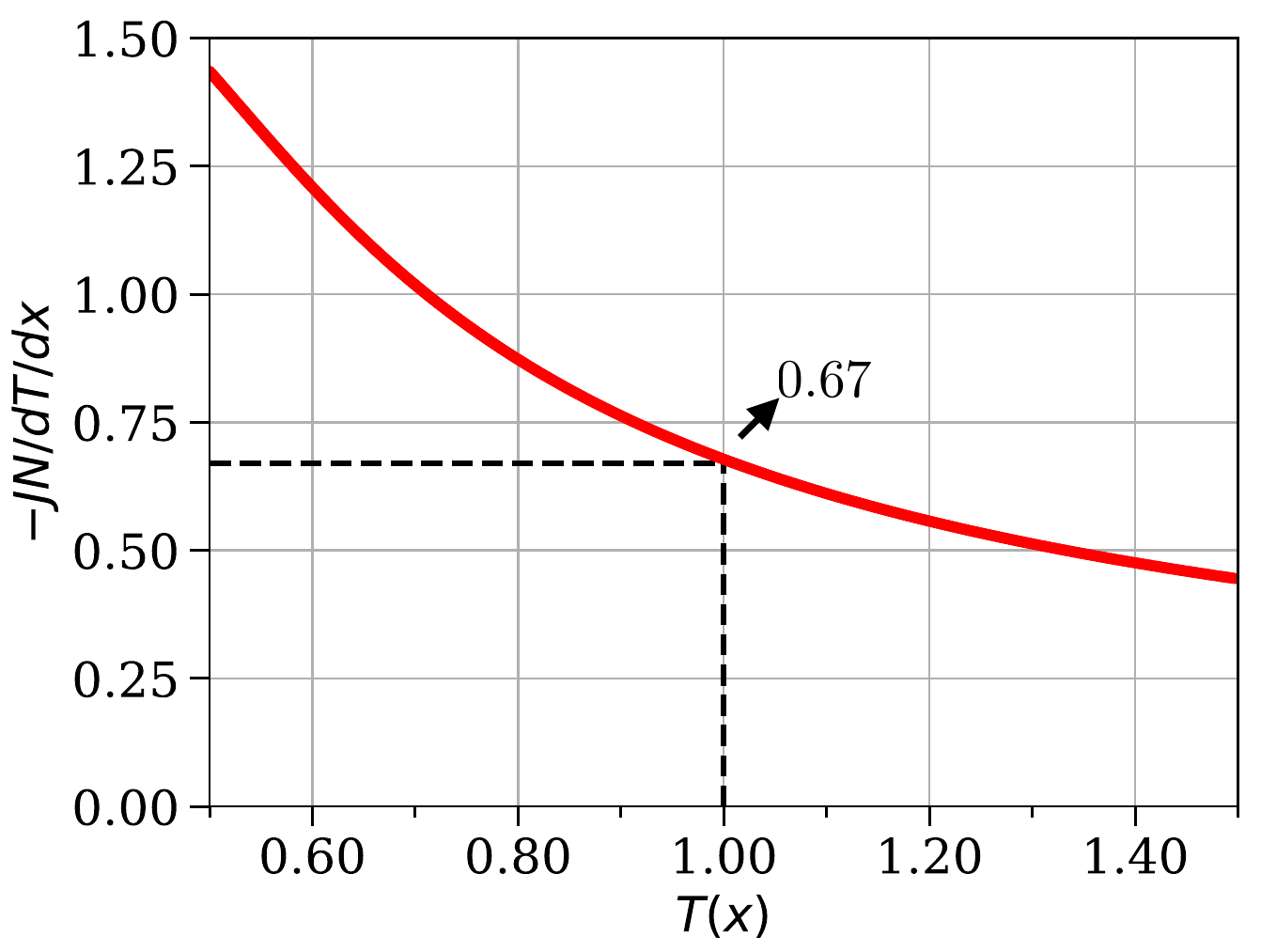}
	\caption{The thermal conductivity ($\kappa = JN/\frac{dT}{dx}$) plotted as a function of temperature [$T(x) \in (0.5,1.5)$] for the quartic pinned Toda chain for system size $N=4096$. The thermal conductivity is a function of temperature in this system. The thermal conductivity at $T=1$ can be seen to be $\kappa \approx  0.67$.}
	\label{fig:kappavsT}
\end{figure}

    \textbf{(b) Quartic pinned Toda chain:}  This corresponds to the case  $z=4$ in Eq.~\ref{eq:open}.
    In Fig.~\ref{quarpin_N} we find that for this case  the destruction of integrability is manifest even at small system size and at low temperatures. The system goes to a NESS which is characterized by a non-linear profile with very small temperature jumps and one also finds that the current scales diffusively as shown in Fig.~ \ref{fig:currscaling}.
    The non-linear profile is due to the temperature-dependent thermal conductivity in the system. The average thermal conductivity [defined as $\kappa = {JN}/{(T_L-T_R)}$] is then independent of system size and has the numerical value $\kappa \approx 0.76$. However the fact that the temperature profile is not linear implies that  the thermal conductivity is a function of temperature and varies significantly within the range $0.5-1.5$. In fact we can find the temperature dependent conductivity  from the local derivative of the temperature profile, thus $\kappa = J /\frac{dT}{di}$.  This is then plotted in Fig.~\ref{fig:kappavsT}, and this gives us   $\kappa \approx 0.67$ for $T=1$.
    
    \section{Simulation results for equilibrium correlation functions}
    \label{sec:Numericalresults:Equilibrium}
    We study the differences in equilibrium correlation functions in the three different setups.
    
    \textit{Details of simulations:}  The system is first equilibrated by attaching Langevin baths at all sites for time $1000$ and then the baths are disconnected and the isolated system is let to evolve while computing necessary observables. The averages for each observable are taken over $~10^5$ initial conditions. The time-step is  $dt \le 0.005$.
    
    \textbf{(a) Quadratically pinned Toda chain:} 
    Consistent with our findings in the NESS, here we find again that the evolution of correlation functions in the harmonically pinned Toda has different behaviour at low and high temperatures. The spatiotemporal spread of energy correlations has distinct scaling at low and high temperatures.
    \\
    
\begin{figure}
        \centering
        \includegraphics[width=0.85\linewidth]{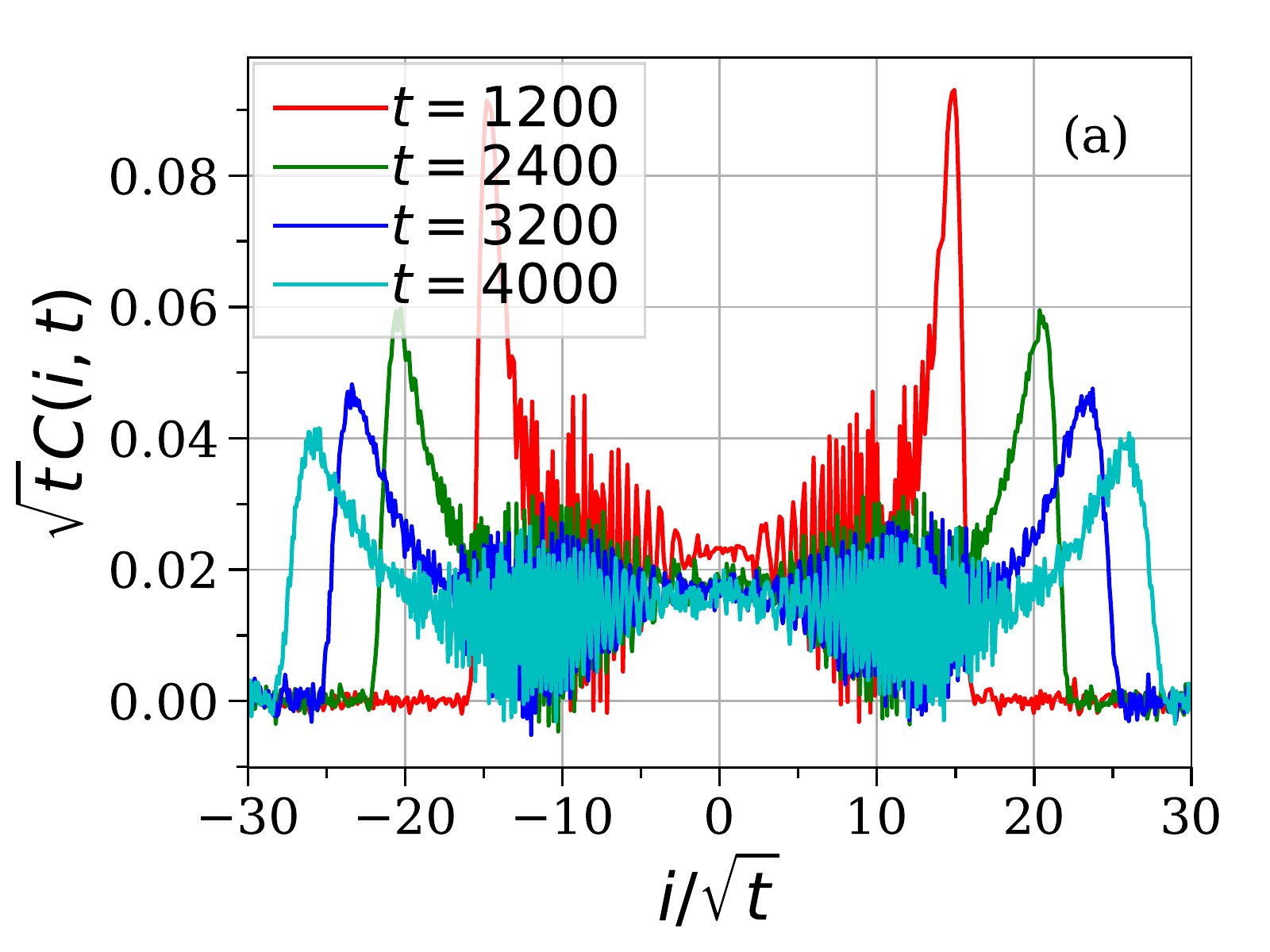}
        \includegraphics[width=0.85\linewidth]{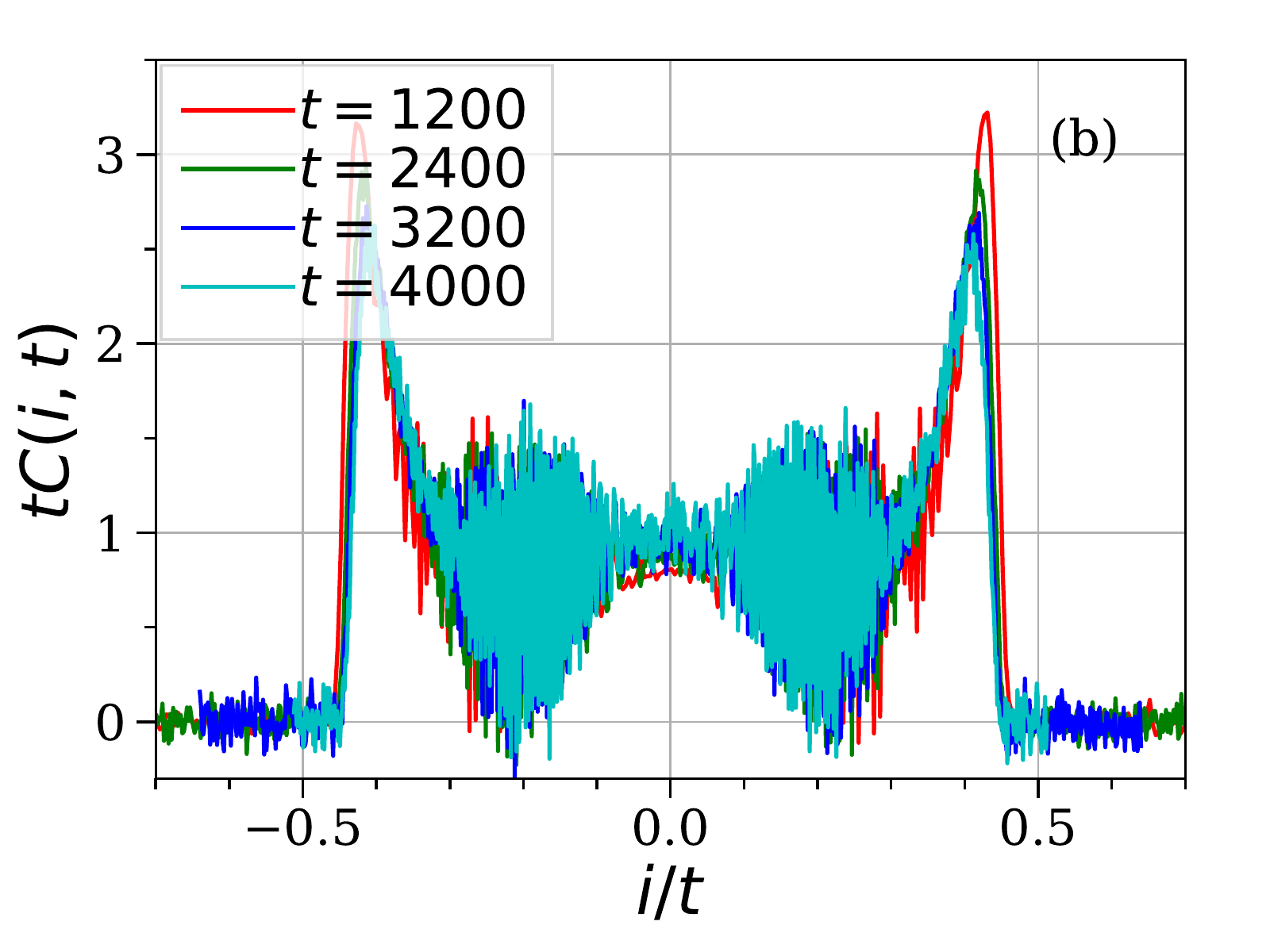}
        \caption{Energy correlations for system size $N=4096$ at low temperature $T=1$ with Toda parameters ($a = b  = 1$, $\nu=2$, $z = 2$) and averaged over $ \approx 10^6$ data points with (a) diffusive scaling of energy correlations. (b) ballistic scaling for energy correlations. We see that at very large times, the equilibrium correlations are still very close to ballistic scaling, although this is not perfect, which is not surprising since we have seen in the non-equilibrium simulations that we start  seeing significant  gradient already at this system size.}
        \label{fig:energycorrlow}
    \end{figure}

\begin{figure}
        \centering
        \includegraphics[width=0.85\linewidth]{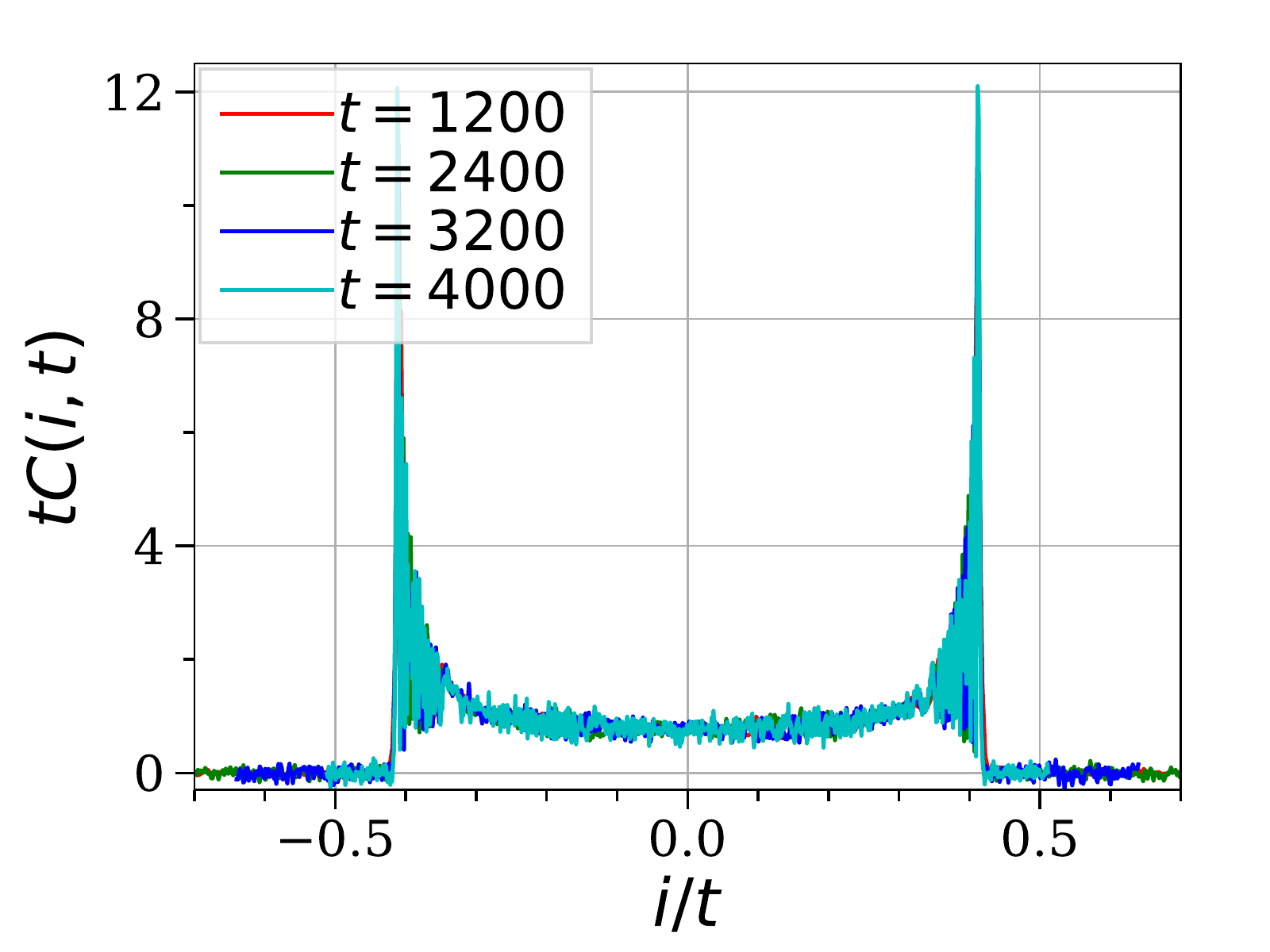}
        
        \caption{Energy correlations for system size $N=4096$ at low temperature $T=1$ for pinned harmonic chain with parameters obtained from an expansion of the Toda potential. As expected we see ballistic scaling  and with strong resemblance with the low-temperature Toda data in Fig.~(\ref{fig:energycorrlow}). The   averaging was over $ \approx 10^6$ data points. }
        \label{fig:energycorrharmonic}
    \end{figure}

    \textit{Low-temperature:}
    At low temperatures, the correlations of the pinned Toda chain is non-diffusive as shown in Fig.~\ref{fig:energycorrlow} and we find an envelope of oscillatory correlations giving a hint that the transport is close to the harmonic case.
    For the overall envelope the scaling of correlations is ballistic. 
    Surprisingly, even at the time when the correlations have reached the boundary, the ballistic peaks still survive along with the oscillatory bulk. This again suggests that at low temperatures the quadratically pinned Toda is close to the integrable harmonic potential at short time and small length scales.  
    In Fig.~\ref{fig:energycorrharmonic} we plot the energy correlations for the harmonic chain where we again see ballistic scaling. For the set of parameters corresponding to the low-temperature harmonic limit of the Toda chain, we see that the ballistic peaks at the ends are in similar positions as that of Fig.~\ref{fig:energycorrlow}(b) and have roughly the same structure. It is expected that at much longer times (requiring much bigger systems), the ballistic peaks will eventually disappear and a diffusive central peak will emerge.  
    
    \begin{figure}
        \centering
        \includegraphics[width=0.85\linewidth]{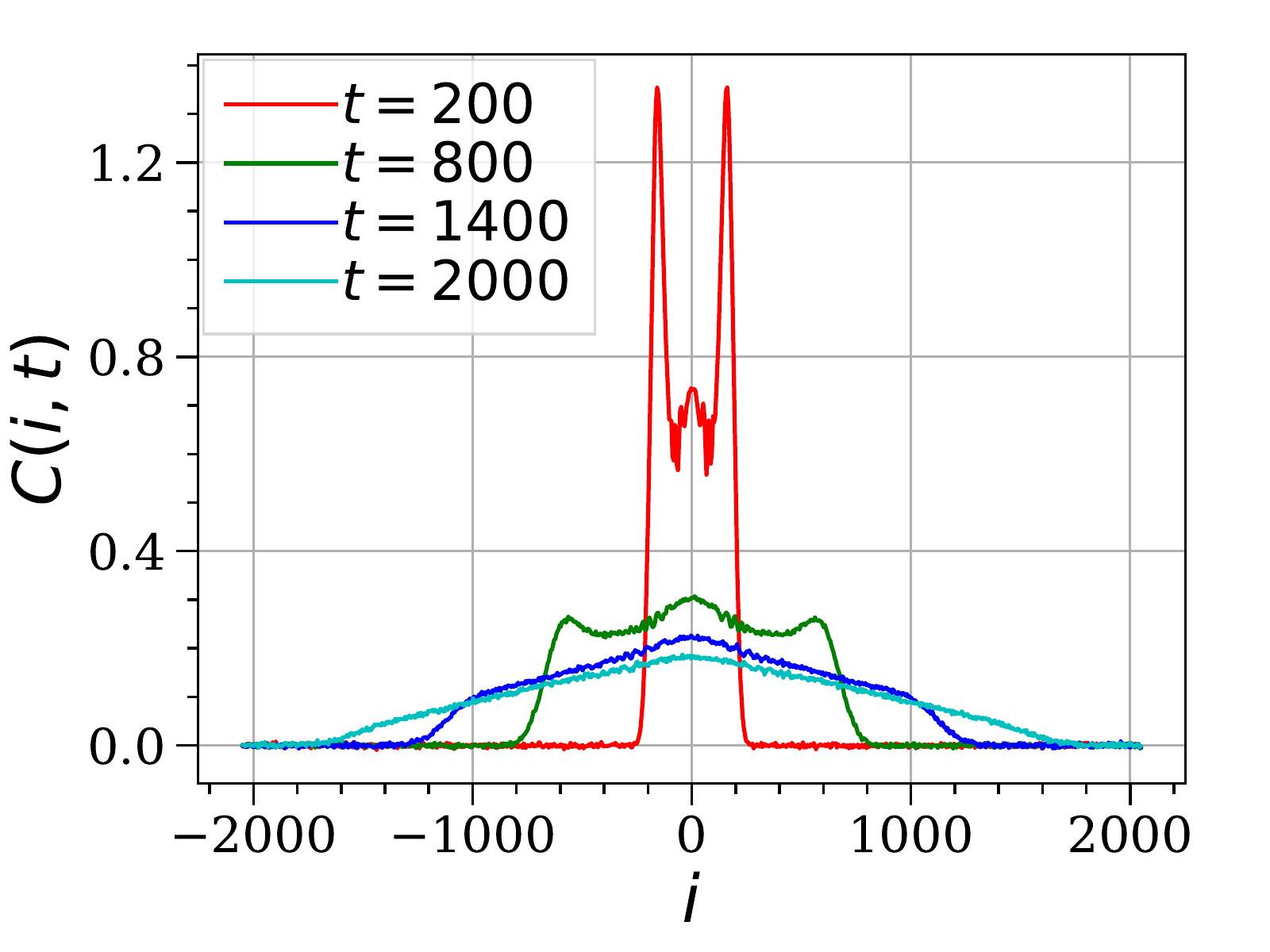}
        \caption{The unscaled correlation functions of energy for Toda with pinning with parameters ($a = b  = 1$, $\nu=2$, $z = 2$). The system is first prepared in equilibrium at temperature $T=20$. The initial times show a ballistic peaks which vanish at late times.}
        \label{fig:corrunscaled}
    \end{figure}
    
\begin{figure}
        \centering
        \includegraphics[width=0.85\linewidth]{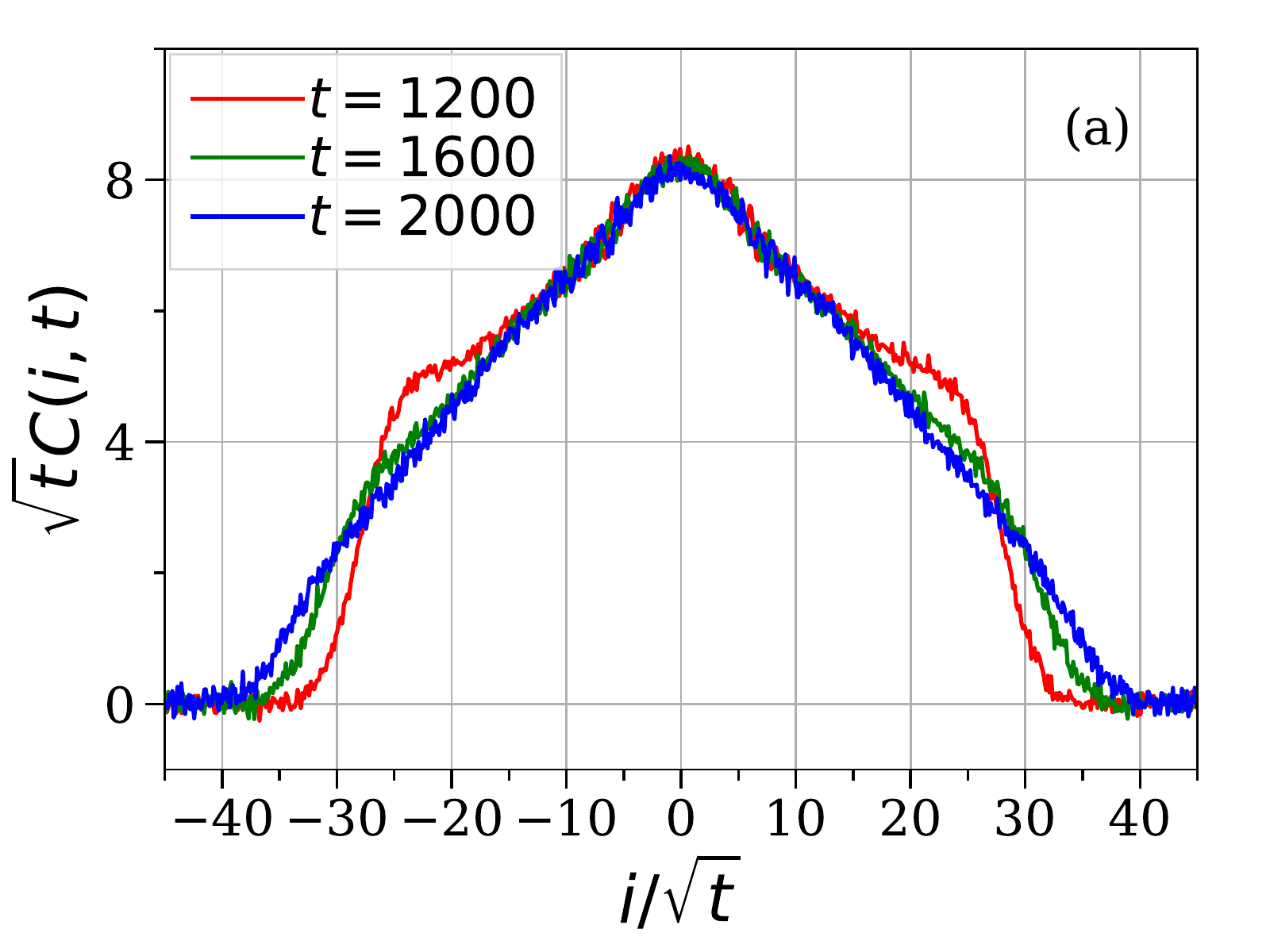}
        \includegraphics[width=0.85\linewidth]{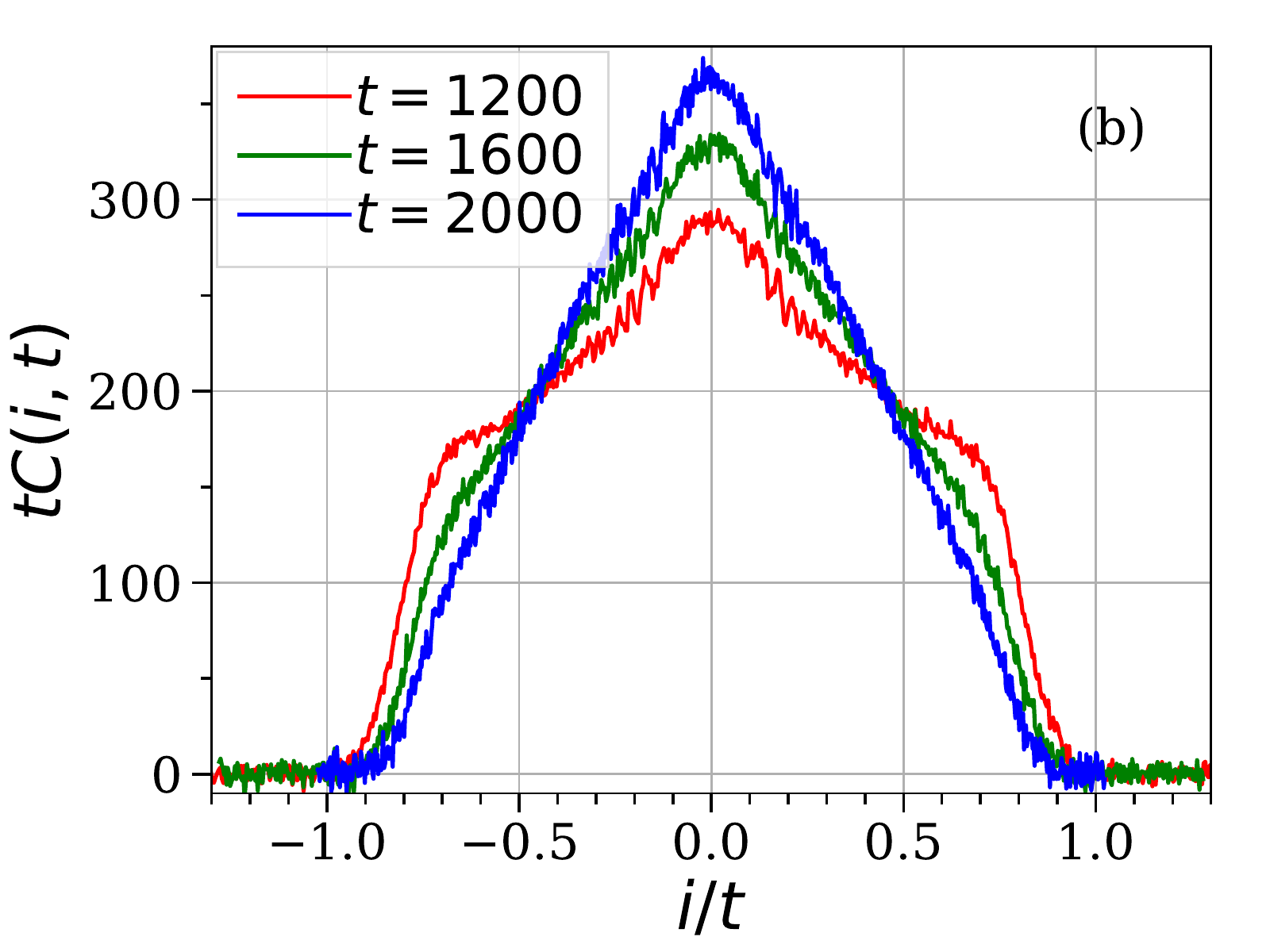}
        \caption{Energy correlations for system size $N=4096$ at high temperature $T=20$ with Toda parameters ($a = b  = 1$, $\nu=2$, $z = 2$) and averaged over $ \approx 10^6$ data points with (a) diffusive scaling of energy correlations, (b) ballistic scaling for energy correlations. We see that there is a better collapse of data with diffusive scaling though we are not yet in the fully diffusive limit. Also note the highly non-Gaussian form.}
        \label{fig:energycorr}
    \end{figure}

    \textit{High temperature:}
    In the high-temperature case, at small times we again see the ballistic peaks in the correlations in Fig.\eqref{fig:corrunscaled} but these now quickly disappear and are replaced by a central peak that scales diffusively. In Fig.~\eqref{fig:energycorr} we see that the spatiotemporal correlations have a much better collapse for diffusive scaling than a ballistic scaling.  However, from the simulated system sizes,  we do not yet see a Gaussian central peak. We also note that the speed of propagation of the ballistic front is twice as that in the low-temperature case.

\begin{figure}
        \centering
        \includegraphics[width=0.85\linewidth]{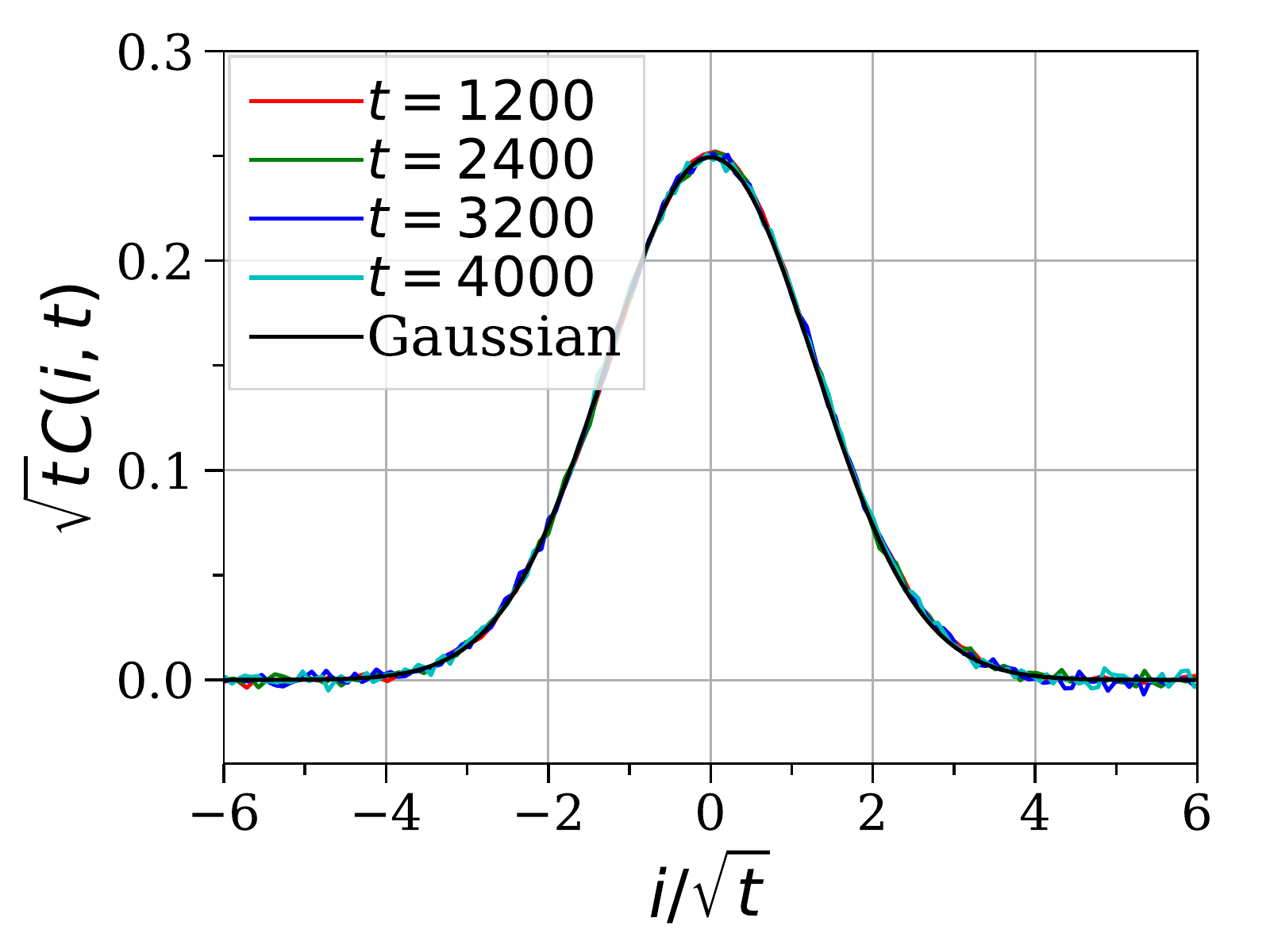}
        
        \caption{Energy correlations for system size $N=4096$ at low temperature $T=1$ with Toda parameters ($a = b  = 1$, $\nu=2$, $z = 4$) and averaged over $ 10^6$ data points. The energy correlations here are purely Gaussian and are very different from the quadratically pinned Toda at both high and low temperatures. The fitted Gaussian has the the form $A {e^{-i^2/4 D t}}/{\sqrt{4 \pi D t}}$, with $A \approx 0.805,D \approx 0.8308$.}
        \label{fig:energycorrquartic}
    \end{figure}

    \textbf{(b) Quartic pinned  Toda chain:}
    For this case, as expected, we find that the equilibrium correlations spread diffusively as shown in Fig.~ \ref{fig:energycorrquartic}. The energy correlations are now seen to be Gaussian, with a diffusion constant $D \approx 0.8308$. This is in sharp contrast to the quadratically pinned Toda chain, where even at high temperatures the spread is non-Gaussian. The thermal conductivity as computed from the equilibrium correlations is given by $\kappa = {D}{C_v}$, where $C_v = N^{-1}\partial \la H \ra/\partial T$ is the specific heat capacity. From numerics, we  compute $C_v \approx 0.83$ at $T=1$. This gives an estimate of $\kappa \approx 0.689$, which is close to the one obtained from non-equilibrium simulations ($\kappa \approx 0.67$) at the same temperature.

    \section{ Numerical results for Lyapunov exponent }
    \label{sec:Numericalresults:Chaos}
\begin{figure*}
        \centering
        \includegraphics[width=\textwidth]{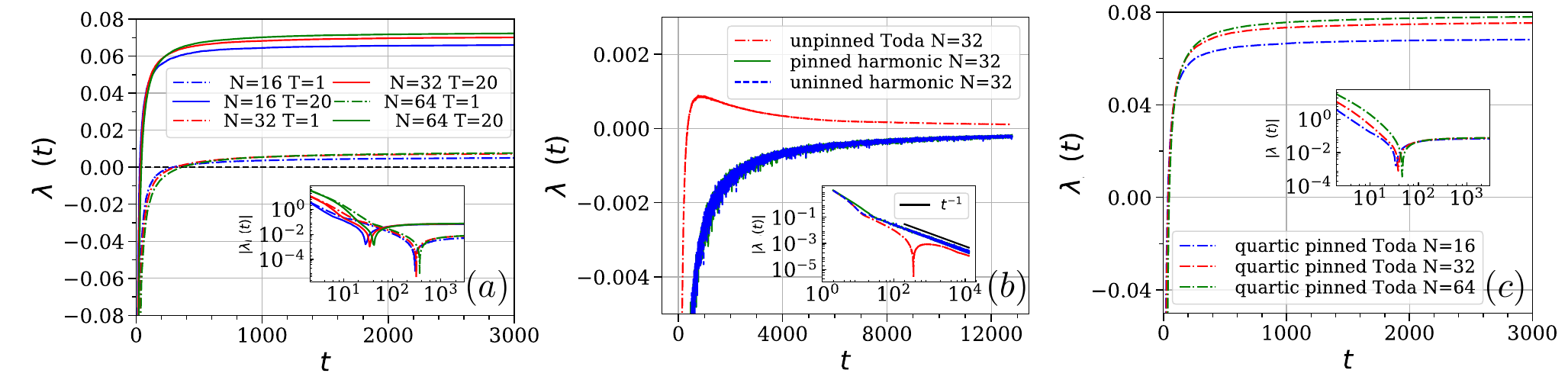}
        \caption{The time-dependent quantity $\lambda(t)$ as defined in Eq.~\ref{eq:lambdaD} is plotted as a function of time for different models. At large times, this quantity gives the largest Lyapunov index in the system. The inset shows the same in a log-log plot. (a) $\lambda(t)$ for the quadratically pinned Toda chain ($a = b  = 1$, $\nu=2$, $z = 2, $) for both low temperatures ( dashed lines: $T=1$) and high temperatures (solid line: $T=20$). At large times both are positive and this shows that the system has positive Lyapunov index and is non-integrable. There is a weak dependence of the Lyapunov index with system size (shown in different colours) for small systems, (b) $\lambda(t)$ for pinned harmonic chain and unpinned Toda chain ($a = b  = 1$, $\nu=0$) which are both integrable. In both the cases, $\lambda(t) \sim 1/t$ for large $t$, (c) $\lambda(t)$ for quartic pinned Toda ($a = b  = 1$, $\nu=2$, $z = 4, T=1$) . Here the Lyapunov index is positive and large even at low temperatures.}
        \label{fig:otoc}
    \end{figure*}

    Finally, we examine the breaking of integrability in the system by studying the Lyapunov exponent $\lambda$ which is a measure of chaos in the system. 
    For integrable models it is known that $\lambda =0$ while for  non-integrable systems we expect $\lambda > 0$. 
    In Fig.~\ref{fig:otoc} we plot  $\lambda(t)$ vs $t$ for the three different cases while the insets  show the same data in log-log scale. 
    
    \textbf{(a) Quadratically pinned Toda chain:} At large times both at low temperature ($T=1$) and high temperature ($T=20$), we see in Fig.~\ref{fig:otoc}(a), the Lyapunov index converge to a finite  positive value. This shows conclusively that the system is non-integrable. However, at low temperatures, the Lyapunov index is much smaller than at high temperatures, which then explains its near-integrable behaviour. 
    In Fig.~\ref{fig:otoc}(b) we plot $\lambda(t)$ for the pinned harmonic chain and unpinned Toda chain ($a = b  = 1$, $\nu=0$) which are both integrable. In both  cases, $\lambda(t) \sim 1/t$ for large $t$ and so we get a vanishing Lyapunov exponent.  Note that for pinned as well as unpinned harmonic chain, $\lambda(t)$ remains negative at all times while for the unpinned Toda $\lambda(t)$ first becomes positive and then decays to zero as $t\to \infty$. 
    
    \textbf{(b) Quartic pinned Toda chain:} In this case, as seen in Fig.~\ref{fig:otoc}(c), the Lyapunov index is again positive and large even at low temperatures which is consistent with the other strong non-integrability signatures in the system. The Lyapunov exponent is approximately $0.078$ whose value is very close to that of the Lyapunov exponent in quadratic pinned Toda in high temperatures. However as we have seen previously, the spatiotemporal correlations in these two models are very different. In the quartic pinned Toda, the equilibrium energy correlations are Gaussian as expected, while for quartic pinned Toda, the correlations are non-Gaussian and the scaling is still imperfect at the system sizes studied. This suggests that the rate of approach to the diffusive transport regime cannot be directly related to the size of the Lyapunov exponent. 
    \section{Conclusions}
    \label{sec:summary}
    Slow relaxation and near-integrable systems have been gaining interest in recent years. Their peculiar properties possibly hold key to a better understanding of chaos and thermalisation. We have studied the slow relaxation in an unusual candidate, the pinned Toda chain where for uniform quadratic pinning, the system shows ballistic like behaviour at low temperatures and small system size and diffusive like behaviour at large temperature or large system sizes. We studied this by looking at equilibrium and non-equilibrium transport properties along with the Lyapunov exponent of the system. We argue that the near-integrable behaviour of the pinned Toda chain arises from the fact that, for small system size and low temperatures, the quadratic pinning leads to the system behaving effectively as a pinned harmonic chain. The anharmonicity appears as a weak perturbation leading to a very large mean free path --- hence the cross-over to diffusive behaviour takes place at very large length and time scales. In contrast, we find that for a Toda chain with quartic pinning, the integrability breaking and the cross-over to diffusive transport is much faster, taking place at small system sizes and low temperatures. For this system we also find that the thermal conductivity obtained from the nonequilibrium  and  equilibrium measurements are in close agreement.

    \section{Acknowledgments}
    We thank C\'edric Bernardin, Stefano Olla, Herbert Spohn, Ovidiu Costin, Rodica Costin and Panayotis Kevrekidis for very useful comments. JAS thanks Mitchell Dorrell for his generous computer programming guidance. The work of JLL was supported by AFOSR grant FA9550-16-1-0037. JAS was supported by a Rutgers University Bevier Fellowship. AD would like to thank the support from the grant EDNHS ANR-14-CE25-0011 of the French National Research Agency (ANR) and from Indo-French Centre for the Promotion of Advanced Research (IFCPAR) under project 5604-2.
    

\begin{thebibliography}{99}
        
        
        \bibitem{Lepri} S. Lepri, ed. ``Thermal transport in low dimensions.'' Springer International Publishing, 2016.
        
        
        \bibitem{BLR} F. Bonetto, J.L. Lebowitz, and L. Rey-Bellet, Fourier’s law: a challenge to theorists, in Mathematical Physics 2000, A. Fokas, A. Grigoryan, T. Kibble, and B. Zegarlinski, eds.,Imperial College Press, London, 2000, pp. 128–150.
        
        \bibitem{LLP2003} S. Lepri, R. Livi, and A. Politi, Thermal conduction in classical low-dimensional lattices, Phys. Rep. {\bf 377}, 1 (2003).
        
        
        \bibitem{dhar2008}   A. Dhar, Heat Transport in low-dimensional systems,
        Adv. Phys. {\bf 57}, 457 (2008).
        
        \bibitem{RLL} Z. Rieder, J. L. Lebowitz and E. Lieb, ``Properties of a harmonic crystal in a stationary nonequilibrium state'', J. Math. Phys. \textbf{8}, 1073 (1967).
        
        \bibitem{spohn} H. Spohn, ``Large Scale Dynamics of Interacting Particles.'' Springer-Verlag, Berlin Heidelberg, 1991.
        
        \bibitem{zotos} X. Zotos, ``Ballistic transport in classical and quantum integrable systems'', J. Low Temp. Phys. \textbf{126}, 1185 (2002).
        
        \bibitem{mazur} P. Mazur, ``Non-ergodicity of phase functions in certain systems'', Physics \textbf{43}, 533 (1969).
        
        \bibitem{suzuki} M. Suzuki, ``Ergodicity, constants of motion, and bounds for susceptibilities,'' Physica \textbf{51}, 277 (1971).
        
        
        
        \bibitem{toda2} M. Toda, ``Solitons and heat conduction,'' Phys. Scr. \textbf{20}, 424 (1979).
        
        \bibitem{shyo} B. S. Shastry and A. P. Young, ``Dynamics of energy transport in a Toda ring'', Phys. Rev. B \textbf{82}, 104306 (2010).
        
        
        
        
        \bibitem{casati2014} S. Chen, J. Wang, G. Casati, and G. Benenti, {\emph Nonintegrability and the Fourier heat conduction law}, Phys. Rev. E {\bf 90}, 032134 (2014).
        
        
        \bibitem{Lepri1997} S. Lepri, R. Livi, and A. Politi, {\emph Heat Conduction in Chains of Nonlinear Oscillators}, Phys. Rev. Lett. {\bf 78}, 1896 (1997).
        
        \bibitem{mai2007} Mai, T., Dhar, A., Narayan, O., {\emph Equilibration and Universal Heat Conduction in Fermi-Pasta-Ulam Chains}, Phys. Rev. Lett. {\bf 98}, 184301 (2007)
        
        \bibitem{Zhao2012} Y. Zhong, Y. Zhang, J. Wang, H. Zhao, {\emph Normal heat conduction in one-dimensional momentum conserving lattices with asymmetric interactions},  Phys. Rev. E {\bf 85}, 060102(R) (2012).
        
        \bibitem{das2013}  S. G. Das, A. Dhar, and O. Narayan, {\emph Heat Conduction in the α−β Fermi–Pasta–Ulam Chain}, J. Stat. Phys. 154, {\bf 204} (2013).
        
        
        \bibitem{hatano} T. Hatano, ``Heat conduction in the diatomic Toda lattice revisited,'' Phys. Rev. E \textbf{59}, R1 (1999).
        
        \bibitem{dhar2001} A. Dhar,  {\it Heat Conduction in a One-Dimensional Gas of Elastically Colliding Particles of Unequal Masses}, Phys. Rev. Lett. {\bf 86}, 3554 (2001).
        
        \bibitem{grass2002} Grassberger, P., Nadler, W., Yang, L., {\emph Heat Conduction and Entropy Production in a One-Dimensional Hard-Particle Gas}, Phys. Rev. Lett. {\bf 89}, 180601 (2002).
        \bibitem{casati} Casati, G., Prosen, T., {\emph  Anomalous heat conduction in a one-dimensional ideal gas}, Phys. Rev. E {\bf 67}, 015203 (2003).
        
        
        
        \bibitem{Narayan2002} Narayan O and Ramaswamy S, "Anomalous Heat Conduction in One-Dimensional Momentum-
        Conserving Systems", Phys. Rev. Lett., {\bf 89} 20 ( 2002).
        
        \bibitem{Beijerin2012} Van Beijeren H, "Exact results for anomalous transport in one-dimensional hamiltonian systems",
        Phys. Revi. Lett., {\bf 108} (18)(2012).
        
        \bibitem{mendl2013} C. B. Mendl and H. Spohn, {\emph Dynamic Correlators of Fermi-Pasta-Ulam Chains and Nonlinear Fluctuating Hydrodynamics},  Phys. Rev. Lett. {\bf 111}, 230601 (2013).
        
        \bibitem{Spohn2013}  Spohn H, "Nonlinear Fluctuating Hydrodynamics for Anharmonic Chains",  J.
        Stat.Phys., {\bf 154} (5) 1191–1227 (2014).
        
        
        
        
        
        \bibitem{doyon2018} B. Doyon, H. Spohn, T. Yoshimura,
        {\emph A geometric viewpoint on generalized hydrodynamics},
        Nuclear Physics B {\bf 926}, 570-583 (2018).
        
        
        \bibitem{kundu2016} A. Kundu and A. Dhar, ``Equilibrium dynamical correlations in the Toda chain and other integrable models,'' Phys. Rev. E \textbf{94}, 062130 (2016). 
        
        \bibitem{prosen2013} T. Prosen and B. Zunkovic,
        {\emph Macroscopic Diffusive Transport in a Microscopically Integrable Hamiltonian System}, Phys. Rev. Lett. {\bf 111}, 040602 (2013).
        
        \bibitem{zhang} Z. Zhang, C. Tang, J. Kang and P. Tong, ``Dynamical energy equipartition of the Toda model with additional on-site potentials,'' Chin. Phys. B \textbf{26}, 100505 (2017).
        
        \bibitem{Cao2018}
        X.~Cao, V.~B.~Bulchandani and J.~E.~Moore,
        ``Incomplete Thermalization from Trap-Induced Integrability Breaking: Lessons from Classical Hard Rods,''
        Phys.\ Rev.\ Lett.\  {\bf 120} no.16,  164101  (2018) .
        
        
        
        \bibitem{jasen2018} J. L. Lebowitz and J. A. Scaramazza, {\emph Ballistic transport in the classical Toda chain with harmonic pinning}, arXiv:1801.07153.
        
        \bibitem{DiCintio2018} Di Cintio, P., Iubini, S., Lepri, S.,  Livi, R.  ``Transport in perturbed classical integrable systems: the pinned Toda chain." Chaos, Solitons  Fractals, 117, 249-254 (2018).
        
        
        
        
        
        
        
        \bibitem{wube} J. Wu and M. Berciu, ``Heat transport in quantum spin chains: Relevance of integrability,'' Phys. Rev. B \textbf{83}, 214416 (2011).
        
        \bibitem{toda} M. Toda, ``Waves in Nonlinear lattice'', Supp. Prog. Theor. Phys. \textbf{45}, 174 (1970).
        
        \bibitem{henon} M. H\'enon, ``Integrals of the Toda lattice'', Phys. Rev. B \textbf{9}, 1921 (1974).
        
        \bibitem{flaschka} H. Flaschka, ``The Toda lattice II: existence of integrals,'' Phys. Rev. B \textbf{9}, 1924 (1974).
        
        
        \bibitem{allen} Allen, Michael P., and Dominic J. Tildesley. Computer simulation of liquids. Oxford university press, 2017.
        
        
        
        \bibitem{rodh} D. Roy and A. Dhar, ``Heat transport in ordered harmonic lattices,'' J. Stat. Phys. \textbf{131}, 535 (2008).
        
        
        
        
        
        
    \end{thebibliography}
\end{document}